\documentclass[preprint,showpacs,prl,preprintnumbers,amsmath,amssymb]{revtex4}

\usepackage{graphicx}
\usepackage{dcolumn}
\usepackage{bm}

\begin{document}
\title{The Origin of Weak Ferromagnetism in CaB$_{6}$}

\author{M. C. Bennett$^{1}$, J. van Lierop$^{1}$, E. M. Berkeley$^{1}$,J. F. Mansfield$^{1}$, C. Henderson$^{1}$, M. C. Aronson$^{1}$,D. P. Young$^{2}$, A. Bianchi$^{2}$, Z. Fisk$^{2}$, F. Balakirev$^{3}$ and A. Lacerda$^{3}$}

\address{$^{1)}$ University of Michigan, Ann Arbor, MI 48109, USA\\
$^{2)}$ National High Magnetic Field Laboratory, Florida State
University, Tallahassee, FL 32310 USA\\
$^{3}$National High Magnetic Field Laboratory, Los Alamos NM 87545
USA}

\begin{abstract}
{We have combined the results of magnetization and Hall effect
measurements to conclude that the ferromagnetic moments of lightly
doped CaB$_{6}$ samples display no systematic variation with
electron doping level. Removal of the surface with acid etching
substantially reduces the measured moment, although the Hall
constant and resistivity are unaffected, indicating that the
ferromagnetism largely resides on and near the sample surface.
Electron microprobe experiments reveal that Fe and Ni are found at
the edges of facets and growth steps, and on other surface
features introduced during growth. Our results indicate that the
weak ferromagnetism previously reported in undoped CaB$_{6}$ is at
least partly extrinsic.}
\end{abstract}
\pacs{75.20.Ck, 75.50.Bb,75.70.Rf} \maketitle


The discovery of ferromagnetism in La doped CaB$_{6}$
~\cite{young} was both remarkable and unexpected. Angle resolved
photoemission (ARPES) measurements~\cite{denlinger2002} find that
undoped CaB$_{6}$ is a direct gap semiconductor, in agreement with
pseudopotential calculations of the electronic structure using the
GW approximation ~\cite{tromp}. Electrons may be introduced by
replacing divalent Ca ions with trivalent La ions, leading to the
formation of a Fermi surface, confirmed by both de Haas - van
Alphen~\cite{hall} and ARPES~\cite{allen} measurements on
Ca$_{1-x}$La$_{x}$B$_{6}$ (0.1$\leq$x$\leq$0.2). Despite the
apparently nonmagnetic and semiconducting character of the system,
ferromagnetism was observed in Ca$_{1-x}$La$_{x}$B$_{6}$ for
(0.001$\leq$x$\leq$0.05)~\cite{young}, with a maximum saturation
moment M$_{S}$=0.07 $\mu_{B}$ per electron for x=x$_{c}$=0.005.
Since these moments would on average be separated by $\sim$ 25
$\AA$ at optimal doping x$_{c}$, it was remarkable that Curie
temperatures as large as 900 K were observed~\cite{ott}.

These observations have been discussed in several different
theoretical frameworks. In the first, ferromagnetism is considered
to be an instability of the low density electron gas which is
introduced by doping, analagous to Wigner crystallization
~\cite{ceperley, ortiz}. While the systematics of how moments and
Curie temperatures depend on electron concentration remain
lacking, the electron concentrations in the La doped samples are
seemingly too large for the high temperature ferromagnetism to be
explained by this scenario~\cite{zong2002}. Subsequently,
attention has shifted to the defect states themselves as the
origin of the moments~\cite{monnier,fisk2002}, which are envisaged
to play a role similar to that of the localized dopant moments in
dilute magnetic semiconductors. However, the central challenge to
both these theoretical scenarios lies with reproducing a Curie
temperature which approaches the Fermi temperature, in a system
with weak and dilute moments.

The possibility that the weak ferromagnetism in CaB$_{6}$ might be
of extrinsic origin has been considered since its discovery. Young
et al~\cite{young} pointed out that the largest moments were found
for the same critical concentration x$_{c}$ in trivalent La, Ce,
and Sm doped CaB$_{6}$ and SrB$_{6}$,  where each dopant ion
introduces a single electron, and that this concentration is
exactly double that found in tetravalent Th doped CaB$_{6}$, where
each Th ion introduces two electrons. These results argue against
accidental contamination of the samples with a ferromagnetic
substance. Subsequently, acid etching
experiments~\cite{otani,mori,taniguchi} found that the
magnetization can sometimes, but not always, be reduced by
removing the surface of the sample. While these experiments would
not necessarily rule out surface contamination, they may also
argue that the ferromagnetism is a feature of the intrinsic
surface electronic structure of CaB$_{6}$, and not of the bulk.
However, recent mass spectrometry measurements~\cite{matsubayashi}
have shown that Fe is present in several ferromagnetic samples, in
rough proportion to the measured saturation moment. Still, the
role of these Fe contaminants in the high T$_{C}$ ferromagnetism
is ambiguous, as they may simply introduce a nucleation center in
the bulk for an extended magnetic state, involving both electrons
from the Fe and from the intentional dopants, much as is found in
dilute magnetic semiconductors~\cite{fisk2002,fisk2002b}.
Explicating the role of these Fe contaminants is crucial for
distinguishing between an essentially intrinsic source for the
high T$_{C}$ ferromagnetism in electron doped CaB$_{6}$, or a
wholly extrinsic mechanism.

The goal of the experiments presented here is to determine whether
an intrinsic factor, electron concentration, or extrinsic factors,
such as ferromagnetic contamination, control the magnitude of the
measured saturation moment in electron doped CaB$_{6}$. We have
performed magnetization and Hall effect measurements on 16
different undoped CaB$_{6}$ single crystals, taken from several
different batches. All samples were single crystals of CaB$_{6}$
prepared using an Al flux technique. None of these crystals were
intentionally doped, and variation in the Ca:B stoichiometry is
presumed to be responsible for the different carrier densities
observed~\cite{fisk2002}. Hall effect measurements were carried
out for temperatures from 0.3 K - 100 K using a 9 Tesla
superconducting magnet and also the 50 Tesla medium pulse length
magnet at the National High Magnetic Field Laboratory in Los
Alamos. In every sample and at every temperature, the Hall voltage
was entirely linear in field, with a slope which indicates that
electrons are the dominant carriers. The electron concentration n
is almost temperature independent, although there is some
indication of magnetic freezeout in the most lightly doped samples
at the lowest temperatures. Magnetization measurements were
carried out using a Quantum Design MPMS magnetometer.

All of the samples we investigated were ferromagnetic, with highly
nonlinear magnetization curves and coercive fields which vary from
$\sim$ 40 - 200 Oe, consistent with previous results on La-doped
CaB$_{6}$~\cite{young}. A Langevin function was fit to each M(H)
curve, and the resulting saturation moment M$_{S}$ at 250 K is
plotted in Fig. 1a for each sample as a function of its electron
density n, measured at 0.3 K. There is no apparent trend in
M$_{S}$ with n, although we note that the moments found in our
samples are in some cases substantially larger than those reported
in Ca$_{1-x}$La$_{x}$B$_{6}$~\cite{young}. The corresponding
moment per electron M$_{S}$/n is plotted in Fig. 1b. The moment
per electron reaches unphysically large values in the most lightly
doped samples, approaching 35 $\mu_{B}$ per electron.
Consequently, we conclude that the variations in the ferromagnetic
moments of our samples are not likely to be controlled by the
measured variations in electron concentration.

Etching experiments indicate that the saturation moment can be
significantly reduced by removing the crystal surface. A 50$\%$
solution of aqua regia was used to etch the samples, and using an
edge profiler we determined that the etch rate for flat surfaces
was 300 $\pm$ 100 $\AA$ per second, approximately linear in time.
The effect of etching on the magnetization of a sample with an
electron concentration of 1.2x10$^{19}$ cm$^{-3}$ is depicted in
Fig. 2a. The first etch removed approximately 6000 $\AA$ of the
surface, and resulted in a 47$\%$ reduction in M$_{S}$. A repeat
of the etch did not result in any further decrease of M$_{S}$,
although it was confirmed that the etch removed an additional 6000
$\AA$ of the sample surface. The temperature dependence of the
resistivity of this sample is plotted before and after the etch in
Fig. 2b. The resistivity is unchanged by the etch, suggesting that
it is representative of the bulk resistivity. The Hall effect was
similarly found to be unaffected by etching. These etching
experiments suggest that at least part of the ferromagnetic moment
is associated with the crystal surface, although they do not rule
out an additional bulk contribution.

Direct observation of the crystal surfaces by electron microscopy
reveals the identity of this inferred surface magnetic phase. We
have performed electron backscattering and microprobe experiments
on the as-grown crystals using a Cameca Microprobe Analyzer with
five wavelength dispersive spectrometers at the University of
Michigan Electron Beam Microanalysis Laboratory. We have examined
a total of 10 different crystals, which overall display two
qualitatively different surface morphologies. A number of our
samples grow with extremely flat and smooth crystal facets, often
with distinct macroscopic terracing. An example of this type of
crystal surface is shown in the electron backscattering image of
Fig. 3a. It is evident from this image that the sample composition
is quite homogeneous, but that the edges of the growth steps are
decorated with a substance with a large atomic number. Microprobe
experiments found that the only atomic species present above
background in this field of view were Ca, B, O, Fe, and Ni.
Microprobe maps for Ca, Fe, and Ni (K-$\alpha$ transitions) are
presented in Fig. 3 b-d. The Ca map(Fig. 3b) indicates that the
surface is clean and unobstructed, except perhaps in the deepest
recesses of the terraces, and that the overall Ca composition is
spatially uniform. Strikingly, Fig. 3c shows that Fe is only found
on the edges of the terrace steps, and on the natural facets of
the crystal. Fig. 3d shows that a small amount of Ni also is found
on these edges, although not everywhere that the Fe is observed.
These decorative phases are likely to be close in composition to
pure Fe and Ni, since we observe no significant enhancement of
other elements in these regions. Some as-grown single crystals
have rough surfaces, although the degree of roughness is highly
variable among samples. Fig 4a is a backscattering image of one
such crystal, showing that the contaminants are trapped in some of
these surface features. The Fe and Ni microprobe maps(Figs 4b,4c)
show that here Fe and Ni are present in roughly equal amounts, but
again are preferentially found in different areas.

Acid etching is generally less effective at removing Fe and Ni
found at crystal facets and growth steps than the Fe and Ni found
trapped in surface features. The Fe microprobe map of Fig. 4d
shows that the surface of this crystal is partially covered by an
Fe film. The Fe film is almost completely removed by a 30 second
acid etch (Fig. 4e), although new areas of Fe decoration are
subsequently revealed at the edges of the crystal facets. We think
it is unlikely that this Fe is actually introduced by the acid
etch. Instead, we suggest that the Fe decoration found at the
facets in this sample extends into the crystal, and that new
decorated facet edges are exposed by etching. A subsequent 60
second etch (Fig. 4f) removed most of the Fe exposed at the facet,
as well as the last remnants of the original film. While the
amount of Fe decorating the facets of the sample in Fig. 4f is
reduced by almost a factor of five by the first 30 second etch
(Fig. 4g), virtually all of the remaining Fe was removed by the
second 60 second etch.  A similar examination of the crystal used
in the magnetization study, shown in the inset of Fig. 2a,
revealed that the Fe remaining after the final etch was similarly
confined to the crystal facets. In every crystal which we
examined, the electron backscattering image performed after
etching  reveals that while the Fe and Ni decoration of the step
edge was removed - or at least greatly reduced - the underlying
steps and facet edges remain and are undecorated.

To summarize, we have found Fe and Ni contaminants preferentially
decorating the edges of facets and step defects in single crystals
of undoped CaB$_{6}$, although in some samples they were also
randomly dispersed over rough surfaces. The contaminants are
generally found within $\sim$ 10,000 - 20,000 $\AA$ of the sample
surface, and can be removed with the surface by acid etching. We
demonstrate that the saturation moment of the sample is
simultaneously reduced by the etching process, although the etch
is more effective at removing Fe and Ni contaminants on the
surface than from the step and facet edges. This observation
suggests that variations in surface morphology from sample to
sample may explain the rather different etching results which have
been reported ~\cite{otani,mori,taniguchi}. Finally, we note that
all the as-grown samples we investigated were to some degree
ferromagnetic, independent of the amount of self-doping introduced
by variations in the stoichiometry of the crystals.

The observation of Fe and Ni at the edges of step defects and
facets introduced during crystal growth indicates that these
contaminants were present in the melt. Step decoration has been
extensively documented as a feature of epitaxial crystal
growth~\cite{brune,speller}. The preferential segregation of
contaminants to the step edges requires that the Fe and Ni ions
have high mobility on the crystal layers as they form, so that
they diffuse to the edges of the layers without trapping. This is
presumably only possible in CaB$_{6}$ because of the very high
degree of crystalline perfection found in the alkali earth
hexaborides~\cite{ott2}. It is unusual, but not unprecedented,
that the minimum energy configuration for the Fe and Ni
contaminants is at the step edge, and not the more highly
coordinated environment found at the bottom of the step. We
suggest that there is considerable lattice strain associated with
replacing a Ca ion with the much larger Fe and Ni ions, and the
step edge is preferred as a nucleation site for groups of defect
ions because the strain can be relaxed here~\cite{klaua}. We
speculate that the polar character of the B$_{6}$ octahedra may
also assist this process by introducing uncompensated negative
charge at the step edge, enabling bonding with the Fe and Ni ions
which would be impossible along a flat surface of B$_{6}$
octahedra. Thus, the Fe and Ni migrate with the step edge, which
moves vertically as the crystal grows layer by layer. In this way,
the Fe and Ni contaminants end up at or near the surface of the
crystal, as we observe.

We acknowledge stimulating discussions with A. H. MacDonald and
his group, and with J. W. Allen, L. M. Sander, and C. Kurdak.
Work at the University of Michigan was performed under the
auspices of the U. S. Department of Energy under grant
DE-FG02-94ER45526. Work at Florida State supported by NEDO. Work
at the NHMFL Los Alamos National Laboratory Facility was performed
under the auspices of the National Science Foundation, the State
of Florida, and the U. S. Department of Energy. EMB was supported
by the University of Michigan NSF-REU program. MCB acknowledges
partial financial support from NHMFL-Los Alamos.

\begin{figure*}[b]
\caption[]{(a): The saturation moment M$_{S}$ at 250 K as a
function of the electron concentration n. Filled circles are from
a previous La doping study ~\cite{young}, and for these four
samples n is deduced from the nominal composition. (b): The same
data, plotted as the moment per electron M$_{S}$/n. The overall
slope results from the normalization to electron concentration n.}
\label{moment_vs_n}
\end{figure*}
\begin{figure*}[b]
\caption[]{(a): The measured magnetization M(H), for an as-grown
crystal($\bullet$), and after 6000 $\AA$ ($\bigcirc$) and 12,000
$\AA$ ($\bigtriangleup$) of the surface have been removed by an
acid etch. Inset: the Fe microprobe map obtained after the second
etch clearly shows residual Fe caught in deep surface recesses.
(b): The temperature dependence of the electrical resistivity
$\rho$ of the as-grown crystal ($\Box$) and after 6000 $\AA$ of
the surface($\bullet$) have been removed.}
\label{effect_of_etching_1}
\end{figure*}
\begin{figure*}[b]
\caption[]{(a): Electron backscattering image of an unetched
single crystal, demonstrating contaminant decoration of the
crystal facets. (b)-(d): Electron microprobe maps of the same
region for Ca (b), Fe (c), and Ni(d).}
\label{Electron_back_scattering_1}
\end{figure*}
\begin{figure*}[b]
\caption[]{(a): Electron back-scattering image of an unetched
single crystal. Electron microprobe maps of the same region for Fe
(b) and Ni (c)  demonstrate that the contaminants are caught in
surface features. Fe microprobe maps of two different unetched
samples (d,g) and after a 30 second acid etch (e,h), and after a
subsequent 60 second etch (f,i).} \label{effect_of_etching_2}
\end{figure*}

\begin{thebibliography}{10}
\bibitem{young}D. P. Young, D. Hall, M. E. Torelli, Z. Fisk, J. D.
Thompson, H. R. Ott, S. B. Oseroff, R. G. Goodrich, and R. Zysler,
Nature $\bf{397}$, 412 (1999).
\bibitem{denlinger2002}J. D. Denlinger, J. A. Clack, J. W. Allen,
G.-H. Gweon, D. M. Poirier, C. G. Olson, J. L. Sarrao, A. D.
Bianchi, and Z. Fisk,Phys. Rev. Lett. $\bf{89}$, 157601 (2002).
\bibitem{tromp}H. J. Tromp, P. van Gelderen, P. J. Kelly, G.
Brocks, and P. A. Bobbert, Phys. Rev. Lett. $\bf{87}$, 016401
(2001).
\bibitem{hall}D. Hall, D. P. Young, Z. Fisk, T. P. Murphy, E. C.
Palm, A. Teklu, and R. G. Goodrich, Phys. Rev. B $\bf{64}$, 233105
(2001).
\bibitem{allen}J. W. Allen, private communication.
\bibitem{ott}H. R. Ott, J. L. Gavilano, B. Ambrosini, P.
vonlanthen, E. Felder, L. Degiorgi, D. P. Young, Z. Fisk, and R.
Zysler, Physica B $\bf{281-282}$, 423 (2000).
\bibitem{ceperley}D. Ceperley, Nature $\bf{397}$, 386 (1999).
\bibitem{ortiz}G. Ortiz, M. Harris, and P. Ballone, Phys. Rev.
Lett. $\bf{82}$, 5317 (1999)
\bibitem{zong2002}F. H. Zong, C. Lin, and D. M. Ceperley,
cond-mat/025339.
\bibitem{monnier}R. Monnier and B. Delley, Phys. Rev. Lett.
$\bf{87}$, 157204 (2001).
\bibitem{fisk2002}Z. Fisk, H. R. Ott, V. Barzykin, and L. P.
Gor'kov, Physica B $\bf{312-313}$, 808 (2002).
\bibitem{otani}S. Otani and T. Mori, J. Phys. Soc.. Japan
$\bf{71}$, 2002.
\bibitem{mori}T. Mori and S. Otani, Solid State Commun.
$\bf{123}$, 287 (2002).
\bibitem{taniguchi}K. Taniguchi, T. Katsufuji, F. Sakai, H. Ueda,
K. Kitazawa, and H. Takagi, Phys. Rev. B $\bf{66}$, 064407 (2002).
\bibitem{matsubayashi} K. Matsubayashi, M. Maki, T. Tsuzuki,
T. Nishioka, and N. K. Sato, Nature $\bf{420}$, 143 (2002)
\bibitem{fisk2002b}D. P. Young, Z. Fisk, J. D. Thompson, H. R. Ott, S. B. Oseroff, R. G.
Goodrich, Nature $\bf{420}$, 144.
\bibitem{terashima} T. Terashima, C. Terakura, Y. Umeda, N.
Kimura, H. Aoki, and S. Kunii, J. Phys. Soc. Japan $\bf{69}$, 2423
(2000).
\bibitem{moriwaka}T. Moriwaka, T. Nishioka, and N. K. Sato, J.
Phys. Soc. Japan $\bf{70}$, 341 (2001).
\bibitem{brune}H. Brune, Surf. Sci. Rep. $\bf{31}$, 125 (1998).
\bibitem{speller}S. Speller, B. Degroote, J. Dekoster, G.
Langouche,J. E. Ortega, and A. Narmann, Surf. Sci. $\bf{405}$,
L542 (1998).
\bibitem{ott2}H. R. Ott, M. Chernikov, E. Felder, L. Degiorgi, E.
G. Moshopoulou, J. L. Sarrao, and Z. Fisk, Z. Phys. B $\bf{102}$,
337 (1997).
\bibitem{klaua}M. Klaua, H. Hoche, H. Jenniches, J. Barthel, and
J. Kirschner, Surf. Sci. $\bf{381}$, 106 (1997).


\end{thebibliography}
\end{document}